\documentclass[12pt,aps,prb,preprint]{revtex4}
\usepackage{epsf}
\usepackage{epsfig}
\usepackage{graphicx}
\usepackage{dcolumn}
\usepackage{bm}
\usepackage{amsmath}
\usepackage{amsfonts}
\usepackage{amssymb}%
\providecommand{\U}[1]{\protect\rule{.1in}{.1in}}
\input epsf
\begin{document}
\title{AC transport in graphene-based Fabry-P\'{e}rot devices}
\author{Claudia G. Rocha}
\affiliation{Institute for Materials Science and Max Bergmann Center of Biomaterials,
Dresden University of Technology, D-01062 Dresden, Germany}
\author{Luis E. F. Foa Torres}
\affiliation{Institute for Materials Science and Max Bergmann Center of Biomaterials,
Dresden University of Technology, D-01062 Dresden, Germany}
\author{Gianaurelio Cuniberti}
\affiliation{Institute for Materials Science and Max Bergmann Center of Biomaterials,
Dresden University of Technology, D-01062 Dresden, Germany}
\date{\today}

\begin{abstract}
We report on a theoretical study of the effects of time-dependent fields on electronic
transport through graphene nanoribbon devices. The Fabry-P\'{e}rot interference pattern
is modified by an ac gating in a way that depends strongly on the shape of the graphene
edges. While for armchair edges the patterns are found to be regular and can be controlled
very efficiently by tuning the ac field, samples with zigzag edges exhibit a much more
complex interference pattern due to their peculiar electronic structure. These studies
highlight the main role played by geometric details of graphene nanoribbons within
the coherent transport regime. We also extend our analysis to noise power response,
identifying under which conditions it is possible to minimize the current fluctuations as
well as exploring scaling properties of noise with length and width of the systems.
\end{abstract}

\maketitle

Time-dependent fields (such as a time-dependent gate voltage or
laser) \cite{Platero2004,Kohler2005,Buettiker2000} allow for novel
electronic transport phenomena beyond the realm of static fields.
Prominent examples include quantum charge pumping
\cite{Thouless1983,Altshuler1999,Switkes1999,Prada2009} and coherent
destruction of  tunneling \cite{Grossmann1991}. Crucial to these
phenomena is the interplay between quantum interference and
photon-assisted processes. An equally relevant role is played by the
electronic structure of the material constituting the device.

\begin{figure}[ptbh]
%\vspace{0.5cm} \center\epsfxsize=8.0cm
%\linewidth\epsfbox{fig1.eps}
\includegraphics[width=6.5cm]{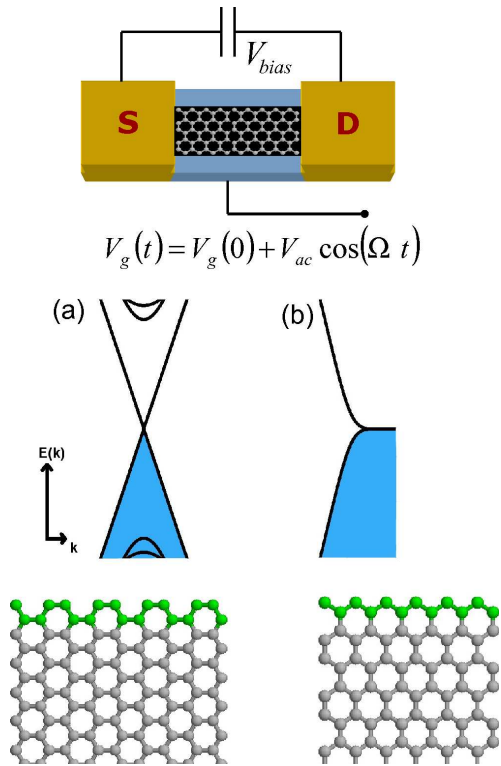}\caption{(color online) Top,
scheme of the device considered in the text. Below, we show the low-energy
dispersion and the atomic structure of (a) an armchair-edge (AGNR) (b) and a
zigzag-edge (ZGNR) graphene nanoribbon. In all calculations we use
$\mathcal{N}=14$ atoms along the width and length of $L=440$~$\mathrm{nm}$ for
the AGNR and $\mathcal{N}=10$ and $L=244\,\mathring{A}$ for the ZGNR case.}%
\label{fig1}%
\end{figure}

Carbon based materials such as carbon nanotubes (CNTs)
\cite{Charlier2007}, graphene \cite{Novoselov1999} and graphene
nanoribbons \cite{Han2007} constitute a promising test ground for
these studies due to their outstanding electrical
properties\cite{Charlier2007,CastroNeto2009} which are at the center
of many promising applications as sensors \cite{Kibis2007}, switches
\cite{delValle2007} and interconnects \cite{Coiffic2007}. Here, our
focus is in graphene nanoribbons, where, thanks to low resistance
contacts, Fabry-P\'{e}rot (FP) quantum interference patterns were
observed \cite{miao2007}. Such low temperature experiments expand
previous studies showing similar phenomena for CNT devices
\cite{Liang2001}. For this last case, besides the conductance
properties, the current noise \cite{Blanter2000} has also been
experimentally probed in the FP regime
\cite{Wu2007,Herrmann2007,YoungKim2007}.

Our contribution complements other recent studies of driven
transport in nanotubes \cite{Orellana2007,Oroszlany2009}, the effects of   
electromagnetic irradiation in both single layer
\cite{Trauzettel2007,Syzranov2008,Rodriguez08,Shafranjuk2008,Oka2009,Rivera2009,Danneau2008}
and bilayer graphene \cite{Shafranjuk2009,Wright2009,Abergel2009}, and quantum pumping in graphene \cite{Prada2009}. In contrast to 2d graphene, in graphene nanoribbons the edges play a decisive role as will be shown later. On the other
hand, studies focusing on AC response of graphene materials usually
resort to a Dirac equation and a linear band approximation,
something that does not always hold for graphene nanoribbons.
Indeed, whenever higher energy bands play an important role or when
the influence of the edges, and/or disorder
\cite{Cresti2008,Long2008} or doping \cite{Biel2009} influences the
electronic structure, these approximations need to be removed.

In this paper we study the effects of ac gating on the conductance
and noise of graphene nanoribbons in the FP regime. In particular,
we show that the interplay between the ac field parameters (field
intensity and frequency) and the typical energy scales of the
ribbon/nanotube (such as level spacing-$\Delta$ and position of van
Hove singularities) can lead to strong modifications on the
conductance and current noise. In contrast to CNTs \cite{Torres2009}
(where the results were independent on the helicity), the shape of
the edges of the graphene nanoribbons turns out to have a dramatic
effect on the interference pattern. Here, two paradigmatic
situations are considered: armchair edges (AGNR) and zizag edges
(ZGNR), see scheme in Fig.\ref{fig1}. For the former, the situation
coincides with the one of CNTs:  the interference patterns observed
in static conditions can be either suppressed, exhibit a revival or
show an ac-intensity independent behavior by tuning the field
intensity and frequency, while the current noise vanishes whenever
the frequency is commensurate with twice the mean level spacing
(\textit{quantum wagon-wheel or stroboscopic effect}). We also extend this
investigation to zigzag-edge nanoribbons and we demonstrate that the
topological shape of the edges strongly determines the behavior of
the patterns.
%For the case
%of ZGNRs the situation is very different mainly due to the presence
%of edge states in the energy spectrum and the intrinsically
%non-linear dispersion. This peculiar electronic structure gives rise
%to more complex Fabry-Perot patterns.
In the following we briefly present the theoretical
framework used for our calculations, then our results and finally
our conclusions.

\section{Methodology}

In this section the general formalism used in our calculations
is outlined. The Hamiltonian of our system is written as:

\begin{equation}
\hat{H} = \hat{H}_{\mathrm{L}} + \hat{H}_\mathrm{R} + \hat{H}_\mathrm{C}+ \hat{H}_\mathrm{T}
\label{hamilt}
\end{equation} where the sublabels L, R and C represent the contributions
from the isolated left, right and central parts, respectively, and T corresponds
to the connection between the leads and the central scattering region.
In the standard tight-binding real space basis
each one of those terms can be written in terms of quantum operators as

\begin{equation}
\hat{H}_{\alpha} = \sum_{i} \epsilon_{i}^{\alpha}\hat{c}_{i}^{\dagger}\hat{c}_{i}^{\phantom\dagger}
+\sum_{\langle ij\rangle} \gamma_{ij}^{\alpha} \hat{c}_{i}^{\dagger}\hat{c}_{j}^{\phantom\dagger}+\mathrm{h.c.},
\end{equation}
where $\hat{c}_{i}^{\dagger}$ ($\hat{c}_{i}$) is the electron
creation (annihilation) operator at site $i$ and
$\alpha=\mathrm{L},\,\mathrm{R}$ or $\mathrm{C}$. The elements
$\epsilon^{\alpha}$ and $\gamma^{\alpha}=\gamma$ are the on-site
energy and the nearest neighbor hopping term, respectively. The
parameter $\gamma =2.7$ eV corresponds to the typical carbon-carbon
hopping element and it is chosen to be our energy unit. The time dependence is introduced by adding a time dependent component to the
on-site energies of atoms located in the scattering region to simulate the
presence of an AC gate plate. Thus, $\epsilon_{j}^{c}=V_{\mathrm{g}}+V_{\mathrm{ac}}\cos(\Omega
t)$ where the AC parameters are $V_{\mathrm{ac}}$, the amplitude of the potential and
$\Omega$, the frequency. The bias voltage is assumed to be equally distributed among the two contacts as required to quasi-ballistic transport and a gate voltage is applied to the central region ($V_{\mathrm{g}}$) which shifts the energy levels.

The contact term is given by

\begin{equation}
\hat{H}_\mathrm{T} = \sum_{\langle ij\rangle} \left \{\gamma_{ij}^{\mathrm{LC}}
\hat{c}_{i}^{\dagger}\hat{c}_{j}^{\phantom\dagger} + \gamma_{ij}^{\mathrm{RC}}
\hat{c}_{i}^{\dagger}\hat{c}_{j}^{\phantom\dagger}\right \} + \mathrm{h.c.}
\label{connection}
\end{equation} and we simulate quasi-transparent coupling between the
electrodes and the sample \cite{Krompiewski2002} using $\gamma^{\mathrm{LC}}=\gamma^{\mathrm{RC}}=\gamma_{\mathrm{t}}=0.7\gamma$. Our  non-interacting model requires
screening by a metallic substrate or by the surrounding gate that lessens
electron-electron interactions. When these interactions come into play effects
beyond our present scope may emerge \cite{Guigou2007,BinheWu2009}.

 %To model the Fabry-Perot setup we consider an infinite
%nanoribbon where the central part of length $L$ is connected through
%weakened hoppings $\gamma_{L(R)}$ to the rest of the sample. In our
%notation, $\gamma=2.7$ eV corresponds to carbon-carbon hopping
%and $\gamma_{L}=\gamma_{R}=\gamma_{t}=0.7\gamma$ is the hopping
%element that simulates the quasi-transparent coupling between the
%electrodes and the sample.
Additional ingredients beyond the stationary theory
have to be considered for the treatment of quantum driven systems. A general
framework valid for noninteracting systems is the use of a Floquet approach \cite{Camalet2003, Kohler2005}, which can also be combined with
Green's function formalism \cite{FoaTorres2005}. Within
this formalism, the DC component of the current can be written as \cite{Kohler2005}%

\begin{equation}
I=\frac{2e^{2}}{h}\sum_{n} \int d\epsilon\left[  T^{(n)}_{\mathrm{RL}}(\epsilon
)f_{L}(\epsilon)-T^{(n)}_{\mathrm{LR}}(\epsilon)f_{R}(\epsilon)\right]  \label{current}%
\end{equation}
being $T^{(n)}_{\mathrm{RL}}(\epsilon)$ the electronic transmission of carriers coming
from right to left leads which might absorb or emit $\mid n\mid$ photons
depending if $n>0$ or $n<0$, respectively. This means that an electron with
initial energy $\epsilon$ has a certain probability of being scattered to a
final energy state of $\epsilon+n\hbar\Omega$.

Assuming an homogeneous driving as well as a weak energy dependence of the self-energy due to the electrodes, both
spatial and time dependencies of the Floquet states can be factorized and the
transport properties can be calculated within simple Tien-Gordon
theory\cite{Tien1963}. Then, the average current over time $t$ can be computed as%

\begin{equation}
\bar{I}=\frac{e}{h}\sum_{n}\mid a_{n}\mid^{2}\int d\epsilon\,T(\epsilon
+n\hbar\Omega)\left[  f_{L}(\epsilon)-f_{R}(\epsilon)\right]
\label{currentTG}%
\end{equation}

\noindent where $T(\epsilon)$ is the transmission in the absence of the driving field. For the case of a harmonic AC
field, the coefficients $a_{n}$ correspond to Bessel functions of the first kind,
$J_{n}(V_{\mathrm{ac}}/\hbar\Omega)$. In turns, the transmission function is written in
terms of Green functions (GF) according to the standard trace formula. On the
other hand, the noise power (zero frequency component of the current-current
correlation function) can be derived for such homogenous driven
system\cite{Kohler2005},

\begin{widetext}
\begin{align}
\overline{S}  &  =\frac{e^{2}}{h}\sum_{n}\int d\varepsilon\left\vert
\sum_{n\prime}a_{n\prime+n}^{\ast}a_{n\prime}^{{}}T(\varepsilon-n\prime
\hbar\Omega)\right\vert
^{2}f_{R}(\varepsilon)\overline{f_{R}}(\varepsilon
+n\hbar\Omega)\\
\nonumber &  +4\Gamma_{L}\Gamma_{R}\left\vert
\sum_{n\prime}a_{n\prime+n}^{\ast
}a_{n\prime}^{{}}G_{1N}^{{}}(\varepsilon-n\prime\hbar\Omega)\left[
2\Gamma_{L}G_{11}^{\ast}(\varepsilon-n\prime\hbar\Omega)-\mathrm{i}\right]
\right\vert ^{2}f_{L}(\varepsilon)\overline{f_{R}}(\varepsilon+n\hbar\Omega)\\
\nonumber &  +\text{same terms replacing (L,1) by (R,N)}%
\label{noiseeq}
\end{align}
\end{widetext}\noindent where $G_{i,j}$ are the retarded Green function between layers $i$ and $j$.
The elements $\Gamma_{L(R)}$ are given in terms of the self-energy of
the corresponding electrode
[$\Gamma_{L(R)}=-\operatorname{Im}(\Sigma_{L(R)})$ with
$\Sigma_{\alpha}=\hat{\gamma}^{\mathrm{C}\alpha}\hat{G}^{\alpha}
\hat{\gamma}^{\alpha\mathrm{C}}$] being $\hat{G}^{\alpha}$ the
retarded surface Green function of the lead $\alpha$.

More general situations where the time-dependent potential is
space-dependent could be solved by using the full Floquet theory
\cite{Moskalets2002, Camalet2003}, methods resorting to equations of
motion \cite{Agarwal2007, Jin2008},  density functional theory
\cite{Kurth2005} or the Keldysh formalism \cite{Pastawski1992,
Jauho1994, Arrachea2005b}. However, even for homogeneous gating, the
approach described above can be heavily demanding in terms of
computational cost depending on the size of the system (matrices of
dimension $\mathcal{N}\times\mathcal{N}$ where $\mathcal{N}$ is the
number of atoms along one layer of the ribbon need to be invert at
each step of the decimation procedure). While this is indeed the
case for the ZGNRs, for AGNRs the problem can be sensibly reduced
thanks to a change of basis transformation which is introduced in
detail in Appendix A.

Based on the tools introduced before we are able to investigate how
the quantum transport properties of GNRs are affected by the AC field. These results are presented in the next session and help to sketch a panorama of  of the response of both armchair and zigzag nanoribbons to such
perturbation. Our numerical analysis and interpretations are also supported by analytical expressions detailed on the appendix.
%Deviations from perfectly homogeneous driving were
%also considered using the full Floquet approach.

\section{\bigskip Results}

%\subsection{ Armchair edge graphene nanoribbons}
\subsection{Direct Current conditions}

As previously mentioned, FP patterns can be generated in electron
waveguide systems with the aid of time independent gate and bias voltages. The
quasi-transparent contacts between the leads and the conductor
confines the electronic wave functions as in light resonant cavity
in which coherent propagation modes can interfere destructively or
constructively generating an interference pattern. In this sense, an
oscillatory behavior of the electronic transmission is observed
while tuning the external voltages and the shape of this pattern is
strictly dependent on the electronic structure of the system.
Depending on the atomic details of their edges, graphene nanoribbons
can reveal remarkable differences on their band structures and
therefore generate distinct patterns as can be seen directly from
Fig. \ref{fig25}. On the upper panel, we present the FP pattern for
an AGNR system and underneath it is shown the picture for a ZGNR.

\begin{figure}[ptbh]
\includegraphics[width=7.0cm]{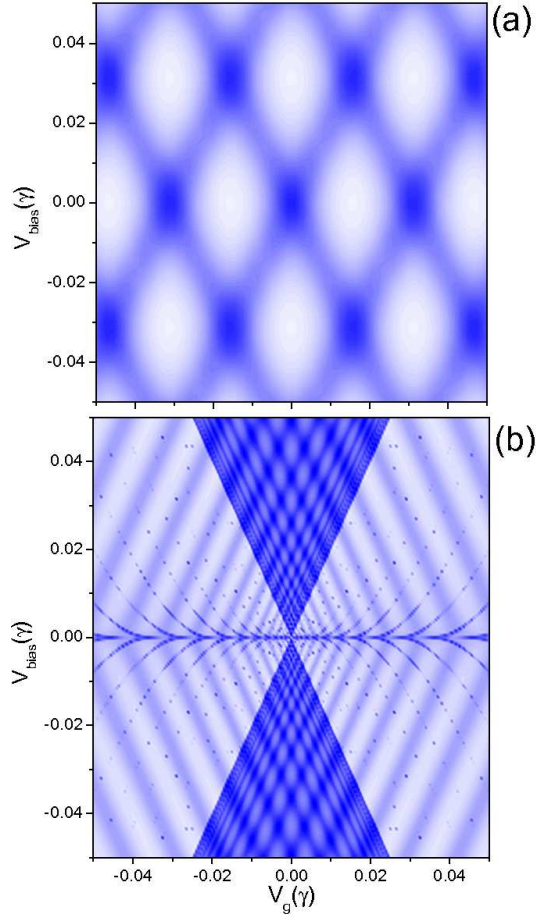} \caption{Fabry-Perot patterns for an (a) AGNR and a (b)
ZGNR. White and dark blue colors correspond to maximum
[$G^{\mathrm{max}}=4e^2/h = G_0$] and minimum conductances
[$G^{\mathrm{min}}\simeq 0.7\,G_0$ for AGNR and
$G^{\mathrm{min}}\simeq 0.2\,G_0$ for ZGNR],
respectively.}%
\label{fig25}%
\end{figure}

Firstly, we focus on Fig \ref{fig25}(a) obtained for an AGNR. The
low energy linear dispersion of AGNRs guarantees a regular energy
spacing level scale, $\Delta$, when the system is brought to
near-perfect ohmic regime. The gate potential shifts the regularly
spaced energy levels while the bias voltage opens the energy window
in which the electronic transmission might take place depending
whether the electronic states interfere destructively or
constructively. Scanning the system through the simultaneously
variation of those two control parameters, the FP panels are drawn
with their characteristic diamonds filling the whole energy range.
The size of the diamonds is given by $\Delta$ and it is possible to
show that $\Delta \simeq (3a_{cc}\gamma/2)\,(\pi/L)$, being
$a_{cc}=1.44$ \AA \,and $L$ the length of the conductor. The
well-defined diamond structures are therefore a manifestation of the
discretization of the linear dispersion of AGNRs and, for this
reason, it is straightforward to infer that metallic carbon
nanotubes also reveal similar patterns. In this sense, we conclude
that such regular behavior is strictly associated with systems
presenting a uniform electronic structure characterized by a well
defined level spacing.

Richer panels as shown in Fig. \ref{fig25}(b) are obtained just
changing the geometry of the nanoribbons to zigzag-edge structures.
It is clear from the schematic band structure shown in Fig.
\ref{fig1}(b) that discretization of the energy levels won't be
regular since the energy dispersion is highly non linear nearby the
flat band. It is only possible to define a characteristic energy
spacing far from the charge neutrality point. From the picture,
three main patterns can be distinguished : (i) one background
oscillation superposed to a (ii) thinner structure restrained in a
cone-shaped and (iii) small emerging lines around
$V_{\mathrm{bias}}=0$ associated with the flat state. The
characteristic FP diamonds are only formed inside the cone
corresponding to the region where $-V/2\leq V_{\mathrm{g}} \leq
V/2$. Due to the non-linear dispersion, the diamonds come in
distinct sizes and evolves to the limit of large level spacing as
the area of the cone increases. We have to mention that the thinner
oscillations revealed mainly in the linear response regime (iii) can
be easily suppressed by temperature since their energy scale is much
smaller than $K_{\mathrm{B}}T$. For ribbons of length
$L=244\,\mathring{A}$ the thin structure could be washed out already
at temperatures of about $\approx \,4\,\mathrm{K}$. Therefore, we
expect to observe FP oscillations only at high bias as observed
experimentally by F. Miao {\it et. al.}\cite{miao2007}. Finally,
although we are coping with the same material, the microscopic
details of the system strongly dictate the shape of the electronic
transmission patterns. We now investigate how these coherent
transport patterns are modified under the presence of AC gate
potentials.

\subsection{Alternate Current conditions}

\begin{figure}[ptbh]
\includegraphics[width=9.0cm]{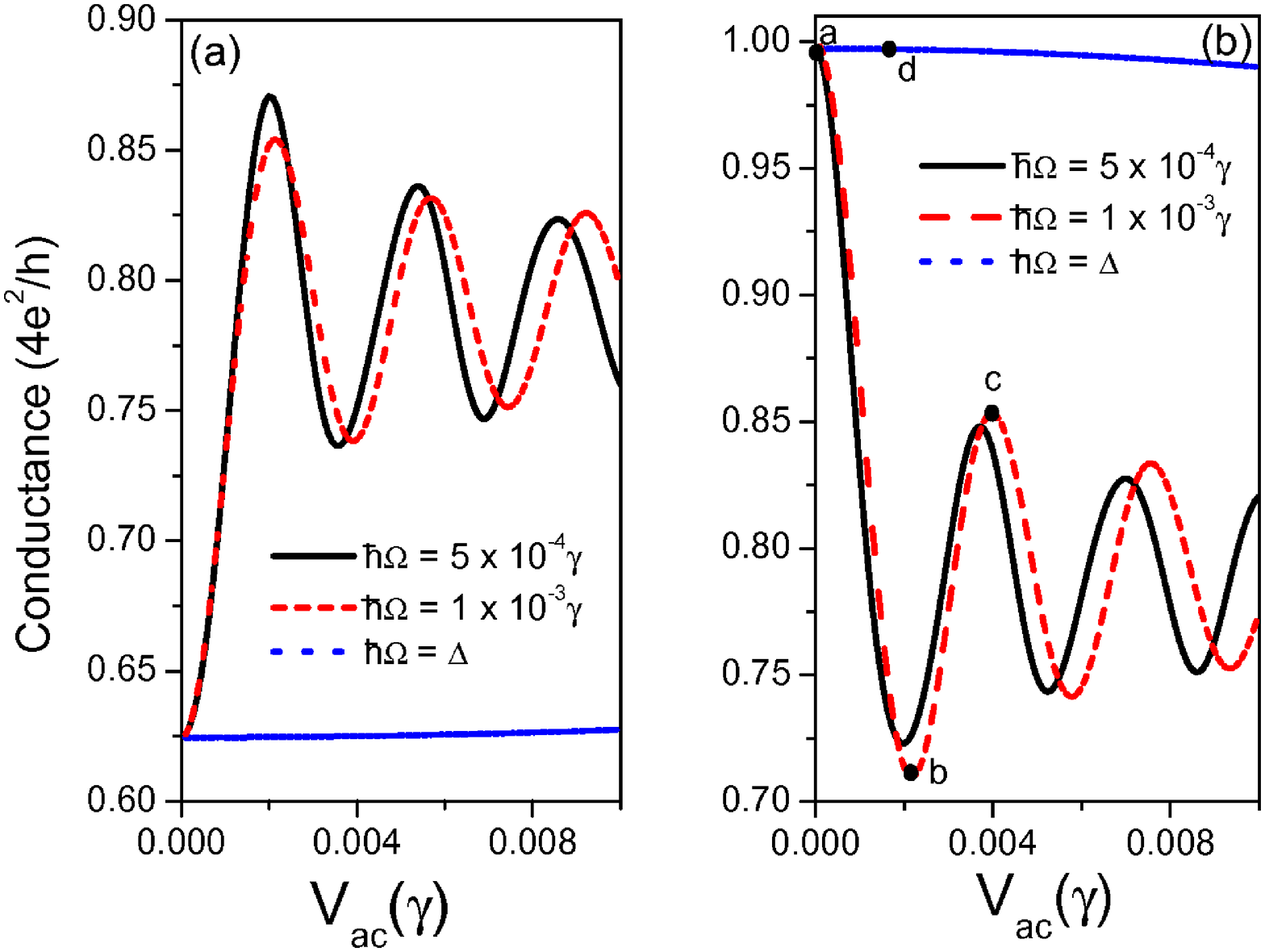} \caption{(color online)
Conductance of AGNR computed as a function of the ac field
intensity. The solid line is for $\hbar\Omega=5.0\times 10^{-4}
\gamma$, the dashed red line for $\hbar\Omega=1.0\times
10^{-3}\gamma$ and the dotted blue line for $\hbar \Omega=\Delta$. Panel (a) corresponds
to $V_g=0$ and in (b) the gate voltage is tuned in such a way that $G_{DC}=1\,G_0$.}%
\label{fig2}%
\end{figure}

As advanced earlier, the FP conductance patterns can be fully controlled with the aid of AC potentials.
Two extra parameters will be used to tune the transport properties
of the ribbons: (i) the amplitude of the AC potential
($V_{\mathrm{ac}}$) and (ii) the frequency ($\Omega$).
%Adding such
%time-dependent perturbation, not only the switching features of the
%nanodevices can be tuned but also the maximum and minimum values of
%the on/off states.
Fig. \ref{fig2} shows the linear conductance at zero bias calculated
for the AGNR structure as a function the intensity of the AC
potential. On the right (left) panel, the conductance of the ribbon
is initially set in a minimum (maximum) value. Different lines
correspond to different frequency values. We can see that the
electronic transmission oscillates and damps to an average value,
$G_{\mathrm{avg}}$, for two of the chosen frequencies. This value
coincides approximately with the conductance in the static situation
(null AC). For $\hbar \Omega=\Delta$, no oscillations in the
transport response is verified, remaining almost constant in the
whole $V_{\mathrm{ac}}$ range. Two completely distinct responses can
be highlighted from this figure as the frequency is changed: an
oscillatory and a constant one (when the frequency matches with the
energy level spacing). The system demonstrates to be strongly
sensitive to frequency variations.

To better understand some of these features, it is effective to appeal to
the adiabatic limit in which $\hbar \Omega$ is the smallest energy-scale
of the system ($\hbar \Omega <<\Delta$). Therefore, the period of
the AC oscillation is long enough so that the system can be
considered instantenously as static. Within this approximation,
the conductance is given by $G=d\bar{I}/dV$ and, from Eq.
\ref{currentTG}, the transmission can be expanded in Taylor series
around $\Omega\rightarrow 0$. Using the identity $\sum_n
J_n^2(z)\exp(-\mathrm{i}n\phi)=J_0[2z\sin(\phi/2)]$ and having in
mind that the transmission function follows a periodic dependence as
$T(\epsilon)=G_{\mathrm{avg}}+A\cos(2\pi \epsilon/\Delta)$, we
obtain

\begin{equation}
G_{\mathrm{ad}}=G_{\mathrm{avg}}+\,J_0\left (\frac{2\pi
V_{\mathrm{ac}}}{\hbar \Omega}\right ) A\cos\left (\frac{2\pi
E}{\Delta}\right ). \label{Gad}
\end{equation}

\noindent It can be noticed that $G_{\mathrm{avg}}$ remains
unaffected by the AC potential while the amplitude $A$ is modulated
by a factor of $J_0(2\pi V_{\mathrm{ac}}/\hbar \Omega)$. Therefore,
the oscillatory contribution is canceled whenever $J_0(z)=0$, i.e.,
the argument $2\pi V_{\mathrm{ac}}/\hbar \Omega$ is a root of $J_0$.
In the following, we investigate how the whole interference patterns
is affected choosing certain values of $V_{\mathrm{ac}}$ and
$\Omega$ which result in special transport conditions on those
curves. For instance, we select the following AC parameters: (a)
$V_{\mathrm{ac}}=0$ (only DC components), first (b) minimum and (c)
maximum conductance and (d) constant transmission on the curve
$\hbar \Omega=\Delta$.

The panels \ref{fig25} show how the full FP interference patterns of
an AGNR interferometer change under the influence of an AC driven
field being its intensity and frequency values marked on Fig.
\ref{fig2} by the letters (a), (b), (c) and (d). The diagram (a)
corresponds to the DC pattern already addressed on the previous
section. These characteristic oscillations are entirely suppressed
when the AC field is driven to the minimum point (b) followed by a
partial revival and a phase inversion when the perturbation is led
to situation (c). Finally, an identical DC FP diagram is recovered
when $\hbar \Omega=m\Delta$, and this reflects in a robustness
feature of the system. Therefore, depending on how the AC parameters
are tuned, it is possible to invert the phase of the oscillations,
to suppress or to recover them. This can be interpreted as a
manifestation of {\it wagon-wheel condition} in the quantum domain
as found for carbon nanotubes \cite{Torres2009}. We now demonstrate
that AGNR also displays such quantum effect.

\begin{figure}[ptbh]
\includegraphics[width=9.0cm]{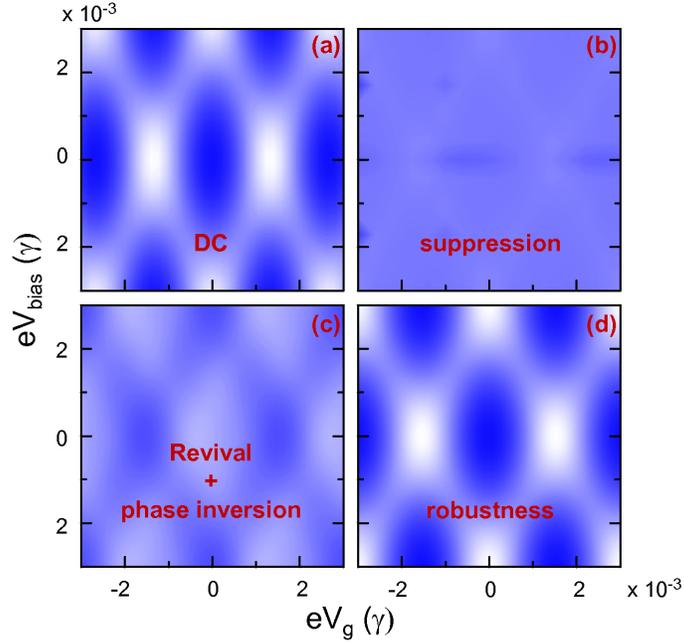} \caption{The panels marked with a, b, c and d,
are Fabry-Perot conductance interference patterns for an AGNR (as a
function of bias and gate voltages) calculated for different driving
frequencies and amplitudes selected in \ref{fig2}(b). White and dark
blue correspond to maximum and
minimum conductances respectively.}%
\label{fig25}%
\end{figure}

\begin{figure}[ptbh]
\includegraphics[width=8.5cm]{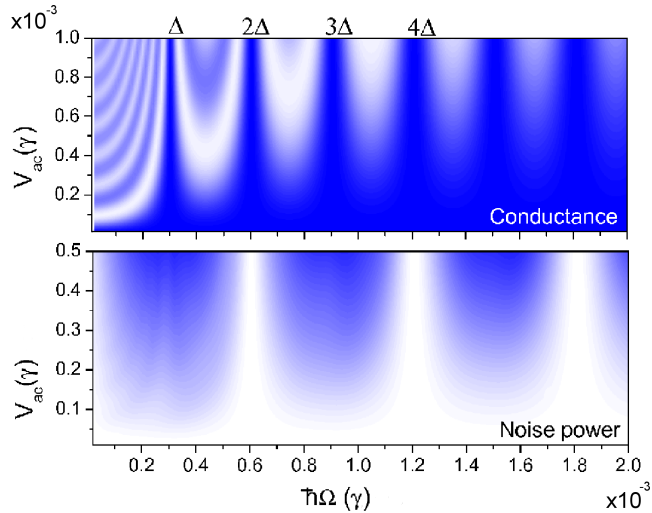} \caption{(color online) Contour
plots showing the conductance (top) and the noise power $\bar{S}$
(bottom) as a function of the driving amplitude and frequency. White
and black colors correspond to
maximum and minimum amplitudes, respectively.}%
\label{fig3}%
\end{figure}

The overall behavior of the FP oscillations in AGNRs is
displayed in the contour plots of Fig. \ref{fig3} together
with noise power results ($\bar{S}$). Blue corresponds to maximum
values of conductance (noise) and white to minimum values. At low
frequency, the results of Fig. \ref{fig3}-upper panel are in agreement with those given by the adiabatic theory and become independent of $\Omega$. On
the other hand, at high frequencies [$\hbar \Omega >\Delta$],
deviations from the adiabatic regime are expected and the
suppression of the oscillations is revealed for higher values of
$V_{\mathrm{ac}}$. Very well defined regions matching with multiples
of the energy spacing level can be also identified from the plots.
As previously stated, whenever $\hbar \Omega=m\Delta$ being $m$ an
integer, the patterns are insensitive to the AC field even under
variations of $V_{\mathrm{ac}}$.

In contrast with the conductance, we find that the noise power (Fig. \ref{fig3}-lower panel) does not behave as in the static
case whenever the whagon-wheel condition is met. In fact, the
suppression of the noise takes place when the frequency is commensurate with an even multiple of the level
spacing. This is a consequence of the fact that the noise
under AC fields is highly sensitive to the phase of the transmission
amplitude which changes by $\pi$ over each resonance. In between
these minima, there are local maxima whose
intensity is proportional to $V_{\mathrm{ac}}$. Summarizing, the noise
recovers the static response whenever $\hbar \Omega=2m \Delta$ being
$m$ an integer. This behaviour is the same as the one found for  and in particular for metallic nanotubes\cite{Torres2009} where this effect
was interpreted as a manifestation of the wagon-wheel condition in
which the static noise behaviour requires a doubling of the
stroboscopic frequency. This occurrence may result in important
technological applications since it is possible to combine high
transmission states with low current noise.

\begin{figure}[ptbh]
\includegraphics[width=10.5cm]{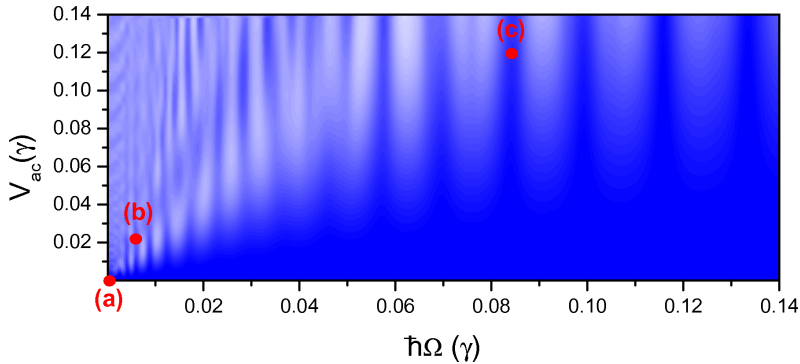} \caption{Contour plot showing
the conductance for a ZGNR as a function of the driving amplitude
and frequency. White and blue colors correspond to maximum and
minimum amplitudes, respectively. The dots ({\bf a}), ({\bf b}) and
({\bf c}) are the set AC parameters pre-determined to obtain
the full FP interference patterns shown in Fig. \ref{fig6}.}%
\label{fig5}%
\end{figure}

\begin{figure}[ptbh]
\includegraphics[width=7cm]{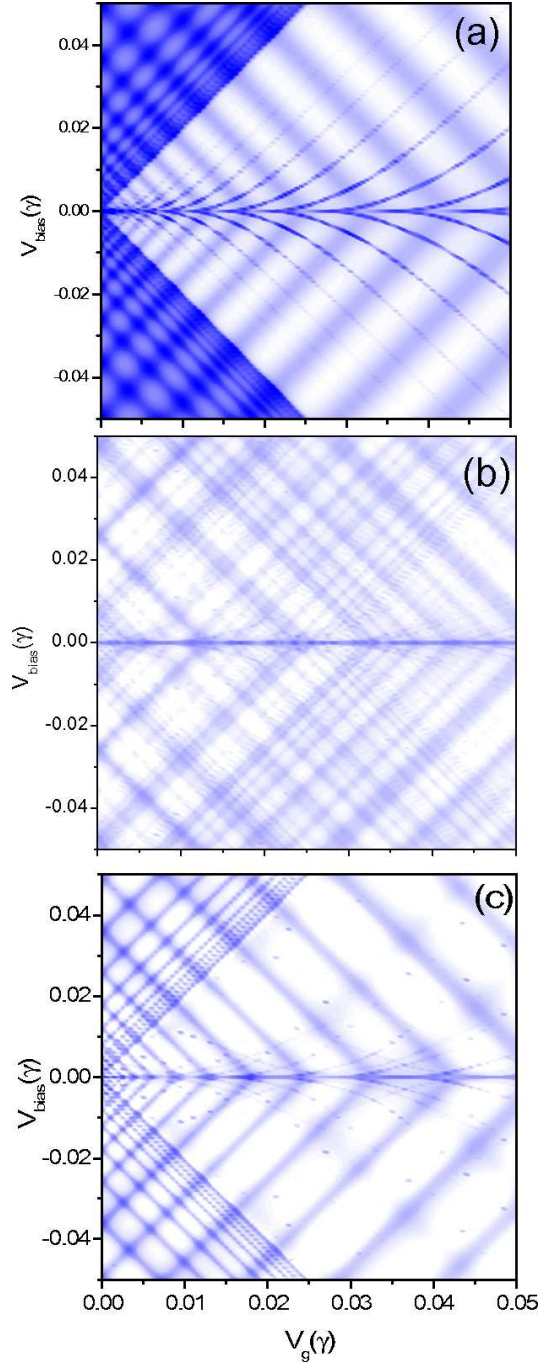} \caption{FP patterns calculated
for a ZGNR at (a) DC and two AC conditions [(b)
$V_{\mathrm{ac}}=0.02\gamma$, $\Omega=0.008\gamma$ and (c)
$V_{\mathrm{ac}}=0.09\gamma$, $\Omega=0.07\gamma$]
marked on Fig. \ref{fig5}.}%
\label{fig6}%
\end{figure}

Under DC conditions, ZGNR already presented richer interference
patterns mainly due to its non-linear dispersion around the flat
state. Combining this highly dense spectrum nearby the Fermi energy
with the application of an external AC driven field, high order
photonic transitions might take place resulting in a even more complicated
pattern, specially in the low frequency range. This can be seen in the AC
contour-plot of Fig. \ref{fig5} which shows the conductance
as a function of the parameters of the driving field. As the
frequency increases, a more regular pattern emerges, showing the same equally spaced segments associated with a
characteristic energy level spacing. A characteristic frequency determining the crossover to the regular FP patterns can
be estimated from the plot. In the following, we present the whole FP-diagrams [Figs.
\ref{fig6}(a) and (b)] when the driving field lies on the particular
points (a) and (b) marked on the contour-plot. The complexity of
diagram \ref{fig6}(a) obtained at adiabatic regime is remarkable
while the robustness is evidenced once more in Fig. \ref{fig6}(b).
These thinner structures can also be eliminated by temperature in a
experimental procedure and for this reason we expect that this
oscillations can only be observed at high frequency limit.

\subsection{Scaling of the current noise with the ribbons length and width}

Geometric aspects such as width and length of the nanoribbon can affect significantly the AC transport properties. For
simplicity, we restricted this investigation to AGNRs due to
their more regular responses and hence they seem to suite better to
applications in electronic nanodevices.
%For instance, the dependence
%of $\overline{S}$ on the nanotube length for two different
%frequencies is shown in Fig. \ref{fig8}.
Increasing the sample length reduces the level spacing and then the
noise is suppressed whenever the wagon-wheel condition is met. We
find that the lower limit for the frequencies required to achieve
this effect is in the order of $10^{2}~\mathrm{GHz}$.
%Fig. \ref{fig8} inset shows the scaling with
%the tube diameter.
As for the scaling with the width, we observed a ten percent increase in the
current noise for widths of about $10$ nm for THz frequencies. This small effect is observed as a result of the onset of a contribution
due to higher subbands. This occurs when inelastic processes can
produce excitations that allow to tunnel over the gap of the
corresponding massive subband, thereby representing a contribution
to the noise from electrons deep in the Fermi sea.

%In contrast, a non-regular behavior is expected for
%ZGNRs which possess a more complex energy dispersion. The scaling
%with the width is expected to be qualitatively the same for both
%systems. Such computationally expensive investigation is beyond our present %capabilities and is left for future works.

\section{Conclusions}

By combining a tight-binding model with a Floquet solution we have
solved for the electronic transport properties of ac-gated graphene
nanoribbons in the Fabry-Perot regime. In contrast to carbon
nanotubes, the interference pattern for nanoribbons depends strongly
on the shape of the edges. For armchair edges, the results coincide
with those obtained for nanotubes and a detailed derivation of both
current and noise properties was presented. The time-dependent field
can be tuned such that the Fabry-Perot oscillations become smoother,
invert their phase, recover the original DC features or even
suppress them. Moreover, whenever $\hbar\Omega$ is an even multiple
of the mean level spacing $\Delta$ it is possible to achieve states
of high conductance with low current noise (quantum wagon-wheel
effect). These calculations for armchair edges greatly benefitted
from a mode decomposition which allows a drastic reduction of the
computational time (see Appendix \ref{apA}). On the other hand, for
zigzag-edges the non-linear energy dispersion nearby the flat band
causes significant changes on the interference patterns making
reasonably easy to distinguish between these two graphene edge
structures. Regular Fabry-Perot oscillations are only visible at
high bias/gate potentials.
%For low bias and gate voltages,
%conductance oscillations occur in a much smaller energy scale and
%for ribbons of length $L=244\,\mathring{A}$ could be suppressed
%already at temperatures of about $\approx \,4\,\mathrm{K}$.

Our work is a first step towards the understanding of the interplay
between quantum interference and ac driving in graphene systems.
Further research aimed at the study of quantum pumping in these
systems is in progress.

\textit{Acknowledgments.} This work was supported by the Alexander
von Humboldt Foundation, by the European Union project
\textquotedblleft Carbon nanotube devices at the quantum
limit\textquotedblright\ (CARDEQ) under contract No. IST-021285-2.
Computing time provided by the ZIH at the Dresden University of
Technology is also acknowledged. We acknowledge Dr. Miriam del Valle
for the valuable discussions.

\section{Appendix A: Eigenchannel/Mode decomposition for graphene nanoribbons}\label{apA}

Solving \textit{brute force} the Hamiltonian to obtain the transport
properties of pristine graphene (armchair edge) nanoribbons, even by using a
decimation procedure, is computationally very expensive. In the following we
detail a way to substantially reduce the calculation size of the problem along the lines of previous work carried out for nanotubes \cite{Mingo2001} and also for graphene \cite{Zhao2009}. The
eigenchannel or mode decomposition scheme proposed below is based on a very
simple idea: rewriting the Hamiltonian in a basis that privileges the
eigenstates in the direction perpendicular to the transport direction.

In an AGNRS, layers of A-type and B-type atoms
alternate along the transport direction. Considering the interaction between
the atoms in these layers, the Hamiltonian can be written in the block form:%

\begin{equation}
H=\left(
\begin{array}
[c]{ccccccc}
&  &  &  &  &  & \\
& E_{1} & V_{1}^{{}} &  &  &  & \\
& V_{1}^{+} & E_{2} & V_{2}^{{}} &  &  & \\
&  & V_{2}^{+} & E_{3} & V_{1}^{+} &  & \\
&  &  & V_{1}^{{}} & E_{4} & V_{2}^{{}} & \\
&  &  &  & V_{2}^{{}} & E_{5} & \\
&  &  &  &  &  &
\end{array}
\right)
\end{equation}
where $E_{i}=\varepsilon_{i}1_{n\times n}$ is the block corresponding to atoms
in the $i-th$ layer and $V_{1}~$and $V_{2}$ are the ones connecting layers of
different type. While $V_{2}$ has a canonical form, $V_{2}=\gamma\,1_{n\times
n}$, the matrix $V_{1}$ can be written as:%

\begin{equation}
V_{2}=\gamma\left(
\begin{array}
[c]{cccc}%
1 & 0 & ... & 0\\
1 & 1 &  & \\
0 & 1 & 1 & ...\\
&  & ... & ...
\end{array}
\right)  . \label{eq-beta2-ACGNR}%
\end{equation}
Note that for the case of carbon nanotubes the top right matrix element,
$[V_{2}]_{1,n}$, is equal to $1$. As argued below, this introduces a major
difference between the (zig-zag) CNT and the (armchair) GNR cases: the lack of
this periodic boundary condition breaks the translational symmetry along the
axis perpendicular to the transport direction.

\textit{The case of zigzag CNTs.} In this case the matrix $V_{2}$ can be
diagonalized, i.e., there is a ($n\times n$) matrix $C$ such that $C^{+}%
V_{2}C$ has a diagonal form. Since $V_{2}$ commutes with $V_{1}$ and $E_{i},$
the change of basis transformation:%
\[
U_{zz-tube}=\left(
\begin{array}
[c]{ccccc}%
... &  &  &  & \\
& C & 0 &  & \\
& 0 & C & 0 & \\
&  & 0 & C & \\
&  &  &  & ...
\end{array}
\right)
\]
gives a block tri-diagonal representation of the Hamiltonian. Each
of these $n$ blocks correspond to an independent mode that can be
represented by a tight-binding chain with alternating hoppings
$\gamma$ and $2\gamma \cos (q\pi/n)$ ($q=0,...,2n$).

\textit{The case of armchair GNRs. }In contrast to the case of zigzag tubes,
the matrix $V_{2}$ [Eq. (\ref{eq-beta2-ACGNR})] cannot be diagonalized.
Therefore, a different strategy is necessary. Inspired by the geometrical
arrangement of the A and B sublattices, an alternative basis transformation
can be adopted:%

\[
U=\left(
\begin{array}
[c]{cccccc}%
... &  &  &  &  & \\
& C_{1} &  &  &  & \\
&  & C_{2} &  &  & \\
&  &  & C_{2} &  & \\
&  &  &  & C_{1} & \\
&  &  &  &  & ...
\end{array}
\right)  ,
\]
where the arrangement of the matrices $C_{1}$ and $C_{2}$ is periodically
repeated with a four layer periodicity (same as the lattice). The matrix
elements of $C_{1}$ and $C_{2}$ are chosen to satisfy hard boundary
conditions:%
\begin{equation}
\left[  C_{1}\right]  _{i,q}=\frac{2}{\sqrt{2n+1}}\sin\left(  \frac{2iq\pi
}{2n+1}\right)  ,
\end{equation}%
\begin{equation}
\left[  C_{2}\right]  _{i,q}=\frac{2}{\sqrt{2n+1}}\sin\left(  \frac
{(2i-1)q\pi}{2n+1}\right)  .
\end{equation}
Interestingly, the blocks of the transformed Hamiltonian $H^{\prime}=U^{+}HU$
are all diagonal. Indeed, the blocks proportional to the identity matrix
remain invariant ($E_{i}^{\prime}=E_{i}^{{}},$ $V_{2}^{\prime}=V_{2}^{{}}$),
while $\left[  V_{1}^{\prime}\right]  _{i,q}=\left[  C_{1}^{+}V_{1}^{{}}%
C_{2}^{{}}\right]  _{i,q}=2\gamma\cos(q\pi/(2n+1)).$ Therefore, the graphene
armchair nanoribbon can be represented as $n$ independent one dimensional
chains with alternating hoppings $\gamma$ and $2\gamma\cos(q\pi/(2n+1))$
being $q=1,...,2n$.

\section{Appendix B: Analytical results for the current noise at the wagon-wheel
condition}

Here we show that for a driven system with a constant level spacing, the
current noise vanishes whenever the frequency is commensurate with twice the
level spacing. We call this modified wagon-wheel or stroboscopic condition,
the \textit{quantum} wagon-wheel condition.

In the following let us consider a single mode. Then, for perfectly
homogeneous driving the noise power (zero frequency component of the
current-current correlation function) can be written according to Eq.
(4)  \cite{Kohler2005}.

In our case, the above expression is the contribution from only one of the
modes in the mode decomposition scheme. However, close to the charge
neutrality point, only two modes contribute. Furthermore, due to symmetry
reasons, their contribution is the same.

Further analysis of the Green functions reveals that for a system with a
constant level spacing the local Green function $G_{11}^{{}}(\varepsilon)$ is
periodic with a period equal to the level spacing $\Delta.$ In contrast,
$G_{1N}^{{}}(\varepsilon)$ (whose phase determines the transmission phase
shift which changes by through each resonance) has a period of $2\Delta$.
These two facts combined with the use of the identity $\sum_{k\prime
}a_{k\prime}^{\ast}a_{k\prime+k}^{{}}=\delta_{k,0}^{{}}$ gives $\overline
{S}(\hbar\Omega=2m\Delta)=0$ as in the static case. It is interesting to
notice that although the conductance at $\hbar\Omega=(2m+1)\Delta$ and
$\hbar\Omega=2m\Delta$ are the same, the noise vanishes only in the
latter case. This gives interesting prospects for achieving maximum
interference\ amplitude with minimum noise in a driven system, much in
consonance with Ref. \cite{strass2005}.

A similar behavior is expected whenever the identity $\sum_{k\prime}%
a_{k\prime}^{\ast}a_{k\prime+k}^{{}}=\delta_{k,0}^{{}}$ is approximately
fulfilled. This is the case for example when $V_{\mathrm{ac}}/\hbar\Omega\ll1$ giving
rise to the features observed in Fig. \ref{fig3} in the vicinity of $\hbar
\Omega=m\Delta$ (odd $m$) and small $V_{\mathrm{ac}}$.

%\bibliographystyle{prsty}
%\bibliography{bibAC}

\end{document}